\documentclass[11pt,a4paper]{article}

\usepackage[dvipdfm]{graphicx}
\usepackage{graphicx,here}
\usepackage{amsfonts,amsmath}
\usepackage{latexsym}
\usepackage{cancel}
\usepackage{jcappub}
\usepackage{color}

\title{Fine Features in the Primordial Power Spectrum}

\author[a]{Kohei Kumazaki}
\author[a]{, Shuichiro Yokoyama}
\author[a,b,c]{and Naoshi Sugiyama}

\affiliation[a]{Department of Physics and Astrophysics, Nagoya University, 
Nagoya 464-8602, Japan}
\affiliation[b]{Institute for the Physics and Mathematics of the Universe (IPMU), 
  The University of Tokyo, Chiba 277-8582, Japan}
\affiliation[c]{Kobayashi-Maskawa Institute for the Origin of Particles and 
  the Universe, Nagoya University, Nagoya 464-8602, Japan}

\emailAdd{kumazaki@a.phys.nagoya-u.ac.jp}
\emailAdd{shu@a.phys.nagoya-u.ac.jp}
\emailAdd{naoshi@a.phys.nagoya-u.ac.jp}

\abstract{ 
A possible origin of the anomalous dip and bump in the primordial power 
spectrum, which are reconstructed from WMAP data corresponding to 
the multipole $\ell=100\sim 140$ by using the inversion method, 
is investigated as a consequence of 
modification of scalar field dynamics in the inflation era.  
Utilizing
an analytic formula to handle higher order corrections to the slow-roll
approximation, we evaluate the relation between a detailed shape of
inflaton potential and a fine structure in the primordial power
spectrum.  We conclude that it is unlikely to generate the observed
dip and bump in the power spectrum by adding any features in the
inflaton potential.  Though we can make a fine enough shape in the
power spectrum by controlling the feature of the potential, the
amplitude of the dip and bump becomes too small in that case.
}

\begin{document}
\maketitle
\flushbottom
\section{Introduction}
Inflation has been the focus of attention as a successful scenario
which gives not only a natural understanding of initial conditions of
the standard Big Bang Cosmology, but also a generation mechanism of
primordial curvature fluctuations, which eventually turn into Cosmic
Microwave Background (CMB) anisotropies and Large Scale Structure of
the universe.  Current precise data of cosmological observations,
e.g., Wilkinson Microwave Anisotropy Prove (WMAP) data
\cite{Komatsu:2010fb}, predict a nearly scale-invariant power spectrum
of primordial curvature fluctuations. As is well-known, such kind of
the power spectrum can be generated from the standard single slow-roll
inflation.  In keeping with the progress of the cosmological
observations, however, small characteristic features in the primordial
power spectrum, e.g., a lack of large-scale power, small dip and
bump at multipole $\ell\approx20\sim40$, a fine oscillating structure
around multipole $\ell\approx100\sim140$ and so
on\cite{Komatsu:2010fb,Larson:2010gs}, have gotten more and more
attention.

In fact, there are many attempts to explain such features in
the primordial power spectrum by considering a non-trivial dynamics
of the scalar field during inflation.  For example, 
several authors\cite{Dvorkin:2009ne,Mortonson:2009qv,Hazra:2010ve}
claimed that the small dip
and bump at multipole $\ell \approx 20 \sim 40$ can be explained due to the
temporarily breaking off from 
the slow-roll dynamics of the inflaton field.
They employed an inflaton potential 
with a sharp step proposed by Adams et al.\cite{Adams:2001vc}:  
\begin{eqnarray}
  V(\phi)&=&\frac12m_\mathrm{eff}^2(\phi)\,\phi^2,\\
  m_\mathrm{eff}^2(\phi)&=&m^2\left[1+ A \,\tanh\left(\frac{\phi - \phi_0}{\Delta \phi_0}\right)\right]. 
  \label{eq:step potential}
\end{eqnarray}
Indeed, the above effective potential can be naturally considered in the two-field 
inflation model with a very small mass difference between two inflaton fields.  
Here the parameter $A$ characterizes the "mass gap".  
For the small value of $A$, the general slow-roll approximation formula proposed by 
ref. \cite{Stewart:2001cd} is applicable to the generation of primordial 
curvature fluctuations.  
Either employing this approximation or numerically solving the perturbation equations, 
the authors obtained a power spectrum of primordial curvature fluctuations 
which can explain the observed small 
dip and bump at $\ell\approx 20\sim40$ in the CMB temperature power spectrum
\cite{Dvorkin:2009ne,Mortonson:2009qv,Hazra:2010ve}.
It should be noted that the width of an oscillation in the power spectrum of 
curvature fluctuations is approximately $\Delta\ln k\sim\,1-1.5$ in order
to match the WMAP data.

Recently, a new oscillating structure around multipole $\ell\approx
100\sim140$ has been reported\cite{Ichiki:2009zz,Ichiki:2009xs}.  The
authors employed the inversion method to reconstruct the primordial
curvature fluctuations from the WMAP data.  The width of the structure
they found in the primordial power spectrum is $\Delta\ln k\sim 0.04$,
which is much finer than the structure at $\ell \approx 20 \sim 40$.
In order to realize such a fine structure in the primordial power
spectrum, Nakashima et al.\cite{Nakashima:2010sa} have proposed a
sudden change of the sound velocity during inflation induced by a
non-canonical kinetic term of the inflaton.  They found fine
oscillations in the primordial power spectrum, while these oscillatory
features continue on the scales smaller than the Hubble radius at the
time of the sound velocity change.  Therefore such a model may screw
up the matching between the observational data and theoretical
predictions at $\ell \gtrsim 200$ which corresponds to the scale of
the first acoustic peak.

In this paper, we investigate other possibilities of generating the
fine structure of the primordial power spectrum by modifying the
inflaton potential.  Employing the general slow-roll approximation
formula, we can easily find the relation between the fine structure of
the primordial power spectrum and the evolution of the slow-roll
parameters during inflation.  Here the dynamics of the inflaton 
determines the evolution of slow-roll parameters.  
For example, it can be shown that 
the mass transition which is described by eq.\eqref{eq:step potential} 
modifies the monotonic evolution of the slow roll parameters and eventually 
generates a structure in the primordial power spectrum.  
We expect that the shorter the mass transition is, the finer the
structure becomes.  

This paper is organized as follows.  In the next section, we briefly
review and investigate an analytical formula to obtain the primordial
power spectrum, that is, the general slow-roll formula.  In section
\ref{primordial_power_spectrum_structure}, based on the formula, we
discuss the possibility of generating fine structures in the
primordial power spectrum due to the modification of the monotonic 
evolution of the slow roll parameters.  The last
section is devoted to the summary and discussion.

\section{General slow-roll formula}

Let us consider a single canonical scalar field $\phi$ whose action is given by
\begin{eqnarray}
S_{\rm matter} = - \int d^4 x \sqrt{\vert {\rm det}g_{\mu \nu}\vert }
\left[
{1 \over 2}
g^{\mu\nu} \partial_\mu \phi \partial_\nu \phi
+V(\phi)
\right]~,
\label{eq:action}
\end{eqnarray}
where $g_{\mu \nu}$ is a metric and $V(\phi)$ represents the potential of the scalar field.
The power spectrum of the curvature perturbation ${\cal R}_c$ induced by quantum fluctuations of 
the scalar field can be defined as 
\begin{eqnarray}
  \langle {\cal R}_c({\bf k}) {\cal R}_c({\bf k}')\rangle 
  = (2\pi)^3 \delta ({\bf k} + {\bf k}'){2\pi^2 \over k^3}{\cal P}_{{\cal R}}(k)~.
\end{eqnarray} 
In the standard single-field slow-roll inflation, the power spectrum is simply 
given by
\begin{eqnarray}
  {\cal P}_{{\cal R}} = \left( {H^2 \over \dot{\phi}} \right)^2_{aH=k}~,
  \label{eq:general power}
\end{eqnarray}
where $H$ is the Hubble parameter, $\dot{\phi}$ is the derivative of the inflaton
with respect to the cosmic time and the subscript
$aH=k$, with $a$ being a scale factor, denotes that the expression is
to be evaluated at the time when the scale of interest exits the horizon.
The above simple formula can be obtained under the assumptions that
the slow-roll parameters are small and keep almost constant values in time.  
Even if the case these assumptions are violated, however, a general formula 
of the power spectrum of primordial curvature perturbations proposed by 
Stewart~\cite{Stewart:2001cd} can be applicable, which is refereed as 
the general slow-role formula.  
Following ref.~\cite{Stewart:2001cd}, 
let us briefly review this formula.

The time evolution of ${\cal R}_{c}(k)$
can be described by~\cite{Stewart:1993bc}
\begin{equation}
  \frac{d^2 v_k}{d\tau^2}+
  \left(k^2-\frac1z\frac{d^2z}{d\tau^2}\right)v_k=0,
\label{eq:evo}
\end{equation}
where $\tau = \int dt/a$ is a conformal time, $v_k \equiv z{\cal R}_c(k)$ and
\begin{equation}
  z\equiv \frac{a\dot\phi}{H}.
\end{equation}
Using the slow-roll parameters $\epsilon$ and $\eta$, 
we have
\begin{eqnarray}
  \frac{1}{z}\frac{d^2z}{d\tau^2}
  =
  2 a^2 H^2 \left[
  1 + \epsilon - {3 \over 2}\eta - {1 \over 2}
  \epsilon \eta + {1 \over 2}\eta^2 
  + {1 \over 2} {\dot{\epsilon} \over H} - {1 \over 2}{\dot{\eta} \over H}
  \right]~, 
  \label{eq:deriz}
\end{eqnarray}
where 
\begin{equation}
  \label{eq:slow roll parameter}
  \epsilon\equiv \frac12\frac{\dot\phi^2}{H^2}\,,\qquad 
  \eta\equiv -\frac{\ddot\phi}{H\dot\phi}\,.
\end{equation}
It should be noted that Eq.\eqref{eq:deriz} is applicable for any values of the slow roll 
parameters, while in the standard slow roll approximation, 
we assume that $\vert \epsilon \vert \ll 1$, $\vert \eta \vert \ll 1$ and their time 
derivatives are also small. 

As far as the slow-roll parameters are order of unity or less, 
we find $(1/z)(d^2z/d\tau^2) = {\cal{O}}(a^2H^2)$.
Hence, on the super-horizon scales ($k \ll a H $),
the above evolution equation~(\ref{eq:evo})
can be approximately reduced as
\begin{eqnarray}
{d^2 v_k \over d\tau^2} - {1 \over z}{d^2 z \over d\tau^2}v_k \simeq 0~.
\end{eqnarray}
From this equation, we can easily find that in the single field inflation
the curvature perturbations ${\cal R}_c$,
remain constant in time on the super-horizon scales.
On the other hand,
on the sub-horizon scales, namely $k \gg a H$,
eq.~(\ref{eq:evo}) can be reduced as
\begin{eqnarray}
{d^2 v_k \over d\tau^2} + k^2 v_k \simeq 0~.
\end{eqnarray}
This equation looks similar to the equation of motion
for the harmonic oscillator in the Minkowski spacetime
and hence the evolution of $v_k$ is decoupled from the background
inflationary dynamics.

From the above discussion, 
the power spectrum of ${\cal R}_c$ strongly
depends on the inflationary dynamics around the horizon crossing time.
Therefore, in order to have a specific feature in the power spectrum, 
the modification of slow-roll dynamics has to take place at the epoch 
of horizon crossing for the corresponding scale.  

Now we are at the position to solve the eq.\eqref{eq:evo} 
by employing the general slow roll approximation. 
Let us first obtain the asymptotic solutions.  
On the super-horizon scales, ${\cal R}_c$ stays constant in time
as we have already shown.  On the sub-horizon scale, we adopt 
the Bunch-Davies vacuum to provide initial conditions.  
Accordingly, we have the asymptotic solutions as 
\begin{equation}
  v_k\to\left\{
    \begin{array}{ll}
      \dfrac{1}{\sqrt{2k}}e^{-ik\tau} \qquad & (k\tau\to-\infty)\\[0.4cm]
      A_kz \qquad & (k\tau\to 0).\\
    \end{array}
  \right.
\label{eq:v_asympt}
\end{equation}
Following ref.~\cite{Stewart:2001cd}, we introduce non-dimensional variables  
$y\equiv \sqrt{2k}v_k$ and $x\equiv -k\tau$.  We also convert $z$ to $f(x)$ as 
\begin{equation}
  f(x) \equiv x z.
\end{equation}
Then equation of motion can be rewritten as
\begin{equation}
  \frac{d^2y}{dx^2}+\left(1-\frac{2}{x^2}\right)y=\frac{1}{x^2}g(x)y
  \label{eq:equation of motion}
\end{equation}
where the {\it source function} $g(x)$ is defined as
\begin{eqnarray}
  g(x) \equiv \frac{1}{f}\left(x^2 \frac{d^2 f}{dx^2} -2 x \frac{df}{dx}\right) .
  \label{eq:definition of source}
\end{eqnarray}
Hence the power spectrum of the comoving curvature perturbation is given by 
\begin{equation}
  {\cal P}_{\cal R}(k)=\left(\frac{k}{2\pi}\right)^2\lim_{x\to 0}\left|\frac{xy}{f}\right|^2,
\end{equation}
Since the asymptotic behavior $x \to \infty$ of the homogeneous solution of 
eq.~(\ref{eq:equation of motion}) has to match with the one $k\tau \to -\infty$ of 
eq.\eqref{eq:v_asympt}, 
we obtain the homogeneous solution of eq.~(\ref{eq:equation of motion}) as
\begin{equation}
  y_0(x)=\left(1+\frac{i}{x}\right)e^{ix}~.
  \label{eq:exact solution}
\end{equation}
Therefore, using the Green's function method,
we obtain the formal solution of eq. ~(\ref{eq:equation of motion})
as
\begin{equation}
  y(x)=y_0(x)+\frac{i}{2}\int_{x}^{\infty}\frac{du}{u^2}g(u)y(u)
  \left[y_0^\ast(u)y_0(x)-y_0^\ast(x)y_0(u)\right].
  \label{eq:solution on green function method}
\end{equation}
Since $g(x)$ can be written in terms of slow-roll parameters and their time derivatives 
as shown below,  we can solve the above formal solution iteratively.   
In fact, under a slow-roll condition; $\epsilon \ll 1$,
we can rewrite the conformal time as
\begin{eqnarray}
x \approx  {k \over aH} {1 \over 1 - \epsilon}~.
\end{eqnarray}
By using this expression, the source function $g$ can be expressed as 
\begin{eqnarray}
  g(x) &=& 2x^2{a^2H^2 \over k^2}
  \left[ 1 + \epsilon - {3 \over 2}\eta - {1 \over 2}\epsilon \eta
  + {1 \over 2}\eta^2 + {1 \over 2}{\dot{\epsilon} \over H} - {1 \over 2}{\dot{\eta} \over H} \right]
  - 2 \nonumber\\
  & \approx &
  {2 \over (1 - \epsilon)^2}\left[ 1 + \epsilon - {3 \over 2}\eta - {1 \over 2}\epsilon \eta
  + {1 \over 2}\eta^2 + {1 \over 2}{\dot{\epsilon} \over H} - {1 \over 2}{\dot{\eta} \over H} \right]
  - 2 \nonumber\\
 & \simeq & -2 \epsilon - 3\eta + {\dot{\epsilon} \over H} - {\dot{\eta} \over H} + {\rm (higher}~{\rm order}~{\rm in}~{\rm slow-roll)}~.
\label{eq:g_in_slow-roll}
  \end{eqnarray}
Unlike the standard slow-roll approximations, we take into account the time derivatives of the parameters, such as 
$\dot{\epsilon}$ and $\dot{\eta}$ in this general slow-roll formula.  

\begin{figure}[t]
  \begin{center}
    \includegraphics[width=0.5\textwidth ,angle = 0]
    {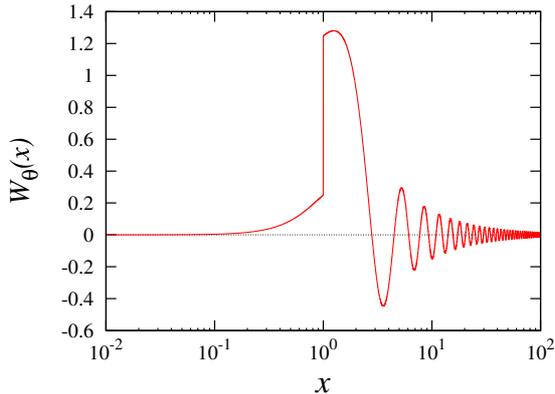}
  \end{center}
  \caption{The window function $W_\theta(x)$.
There is a sharp cut-off at $x=1$ and oscillatory behavior is shown at $x\gg 1$.  }
  \label{fig:window function}
\end{figure}

Then, up to the first order in the source function $g$,
we develop an analytical formula of the power spectrum of the curvature perturbations as
\begin{equation}
  {\cal P}_{\cal R}(k)
  =\left(\frac{k}{2\pi}\right)^2\frac{1}{f_\star^2}\left[1+\frac23
    \left(\frac{1}{f}\frac{d f}{dx}\right)_\star
   +\frac23\int_0^\infty\frac{du}{u}
    W_\theta(u)g(u)+\mathcal{O}(g^2)\right],
  \label{eq:analytical formula of power spectrum}
\end{equation}
where $\star$ denotes the value at the horizon crossing time, $x_\star=-
k\tau_\star=1$, and the {\it window function} 
\begin{equation}
  \label{eq:step window function}
  W_\theta(x) \equiv W(x)-\theta(x_\star-x), 
\end{equation}
with 
\begin{eqnarray}
  \label{eq:window function}
  W(x)&\equiv &\frac{3\sin{(2x)}}{2x^3}-\frac{3\cos{(2x)}}{x^2}
  -\frac{3\sin{(2x)}}{2x}, \\
  \label{eq:step function}
  \theta(x)&=&\left\{
    \begin{array}{ll}
      1 &\qquad (0<x)\\
      0 &\qquad (0>x).\\
    \end{array}
  \right.
\end{eqnarray}
Figure~\ref{fig:window function} shows the form of the window function.
In this figure, since $x=1$ is the horizon crossing time for a given wavenumber,
$x \ll 1$ corresponds to the time after the horizon exit (on super-horizon scales)
and while $x \gg 1$ to the time before the horizon exit (on sub-horizon scales).
This function reflects the behavior of the homogeneous solution $y_0(x)$.
Due to the oscillatory behavior of the window function 
for $x \gg 1$
and the sharp cut-off for $x \ll 1$ coming from the step function,
we can easily find that the integral term of the formula mostly
depends on the behavior of the source function around the horizon cross time.
Therefore it is clear that the specific feature on the power spectrum induced by 
the modification of slow-roll dynamics appears only at the scales 
crossing the horizon when the modification takes place.  

If slow-roll conditions are satisfied, the source function $g$ and $(df/dx)/f$
are approximately zero. Then, we verify that  eq. 
\eqref{eq:analytical formula of power spectrum} 
leads to the well-known slow-roll expression~\eqref{eq:general power};
\begin{eqnarray}
  {\cal P}_{\cal R}(k)=\left(\frac{k}{2\pi}\right)^2\frac{1}{f_\star^2} 
  =\left(\frac{H^2}{\dot\phi}\right)^2_{aH=k}~.\nonumber
\end{eqnarray}
Therefore, we can conclude that eq. \eqref{eq:analytical formula of power spectrum}
is a natural extension of the slow-roll approximation to the generalized one.  
In appendix~\ref{sec:valid},
we check the validity of this general slow-roll formula.

\section{Can we make fine structures in the primordial power spectrum?}
\label{primordial_power_spectrum_structure}

In the previous section, we have briefly reviewed the general
slow-roll formula to evaluate the primordial power spectrum in which
we iteratively take into account the modification of the slow-roll
dynamics.  The analytical formula eq. (\ref{eq:analytical formula of
power spectrum}) consists of the window function $W_\theta(x)$ which
reflects the behavior of the homogeneous solution $y_0(x)$, and the
source function $g(x)$ related to the inflationary background dynamics
through the slow-roll parameters.  Accordingly, the fine structures of
the primordial power spectrum depends on the form of the window
function and also that of the source function.  In this section, using
the analytical formula, we study the modification of the primordial
power spectrum by considering the effect of the modification of the
slow-roll dynamics.  We focus on the relation between the fine
structures of the primordial power spectrum and the form of the source
function $g$ with keeping the form of the window function. 

\subsection{Source function $g$}

\begin{figure}[t]
  \begin{tabular}{cc}
    \begin{minipage}{0.5\textwidth}
      \begin{center}
        \includegraphics[width=\textwidth]
        {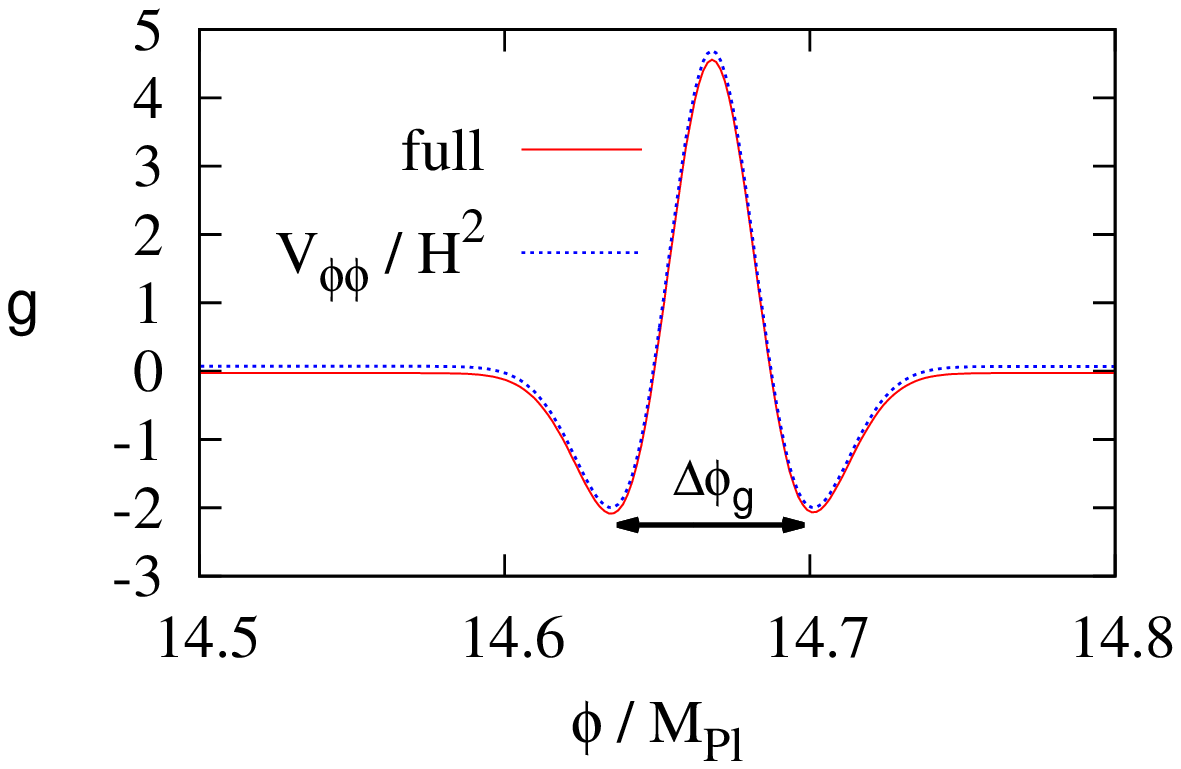}
        \caption{The source functions $g(\phi)$ with 
    all terms of the right hand side of eq. \eqref{eq:definition of source}
                 (red solid line) and with only 
          $V_{\phi\phi}/H^2$ (blue dotted line) for the bump model.}
        \label{fig:source and pote} 
      \end{center}
    \end{minipage}
    \begin{minipage}{0.5\textwidth}
      \begin{center}
        \includegraphics[width=\textwidth]
        {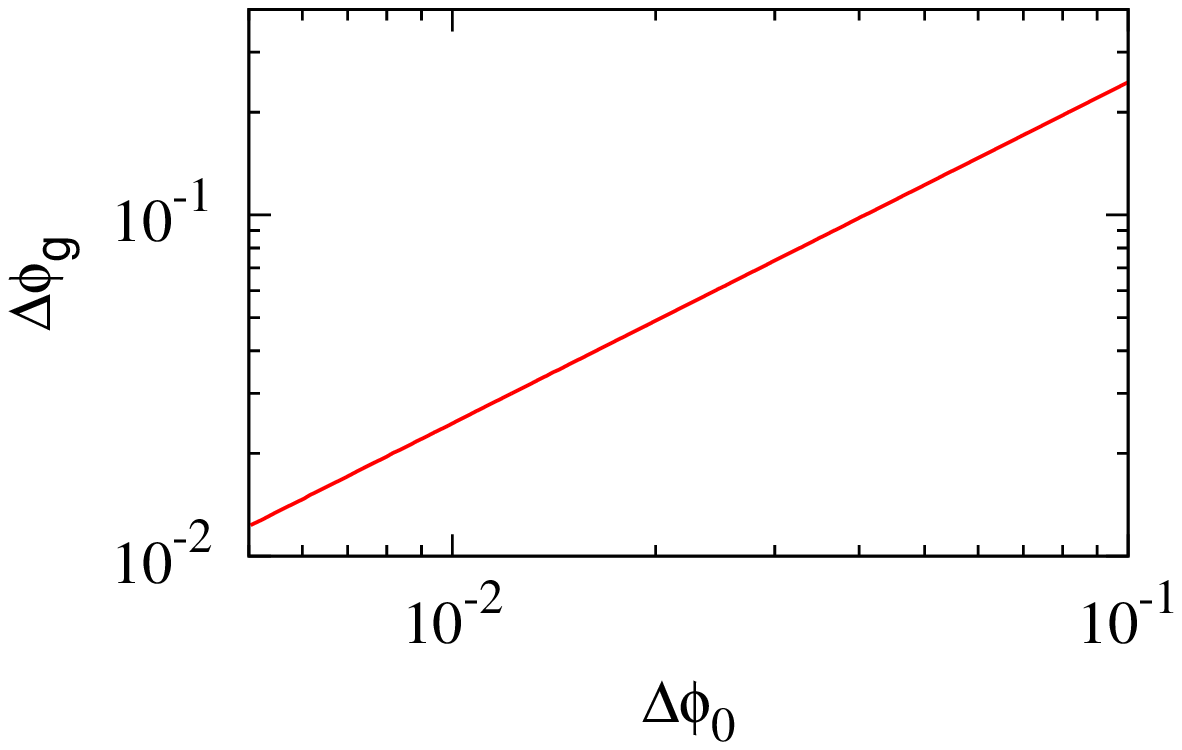}
      \caption{The relation between the width of the inflaton potential $\Delta\phi_0$ 
and the one of the source function $\Delta\phi_g$ for the bump model. }
      \label{fig:width compare} 
      \end{center}
    \end{minipage}
  \end{tabular}
\end{figure}

First,
let us briefly mention how the functional form of $g$
can be related with theoretical models of inflation.
Stewart \cite{Stewart:2001cd} has calculated the 
primordial power spectrum 
by making use of the general slow-roll formula with changing
the functional form of $f(x)$. 
In this paper, however, we focus on the form of the source function
$g(x)$ instead of $f(x)$ which we can construct from $g$
by using eq. \eqref{eq:definition of source}.

In eq.~(\ref{eq:g_in_slow-roll}), we gave the expression for $g$ with respect to the slow-roll parameters
defined by eq.~(\ref{eq:slow roll parameter}).
Here, we give another expression for $g$
in terms of derivatives of potential $V(\phi)$ such as 
$V_{\phi} \equiv dV/d\phi$ and $V_{\phi\phi} \equiv d^2V/d\phi^2$, 
time derivative of $\phi$ and 
Hubble parameter $H$ as
\begin{eqnarray}
  g&=&aH\tau\,\left[\,\frac32\frac{\dot\phi^2}{H^2}
    +\frac{\dot\phi^4}{2H^4}
    +\frac{V_{\phi\phi}}{H^2}
    +\frac{\dot\phi\,V_\phi}{2H^3}\,\right]~.
\end{eqnarray}
In order to investigate which term of the right hand side of this equation mainly
contributes to $g$,
in figure \ref{fig:source and pote} 
we show
the source function $g$ and 
$V_{\phi\phi}/H^2$ for a bump model whose potential is given by
\begin{eqnarray}
V &=& {1 \over 2}m_{\rm eff}^2(\phi) \phi^2~, \nonumber\\
m_{\rm eff}^2 &=& m^2 \left[1 + c \exp \left( - (\phi - \phi_0 )^2 / \Delta \phi_0^2 \right) \right]~.
\label{eq:bumppotential}
\end{eqnarray} 
From this figure, it is found that the form of $g$ is dominated by 
$V_{\phi\phi}/H^2$.  
We find that this is also true for the case of a step in the potential.  
Therefore we can conclude that $V_{\phi\phi}$ controls $g(\phi)$ when the dynamics of the inflaton 
modifies due to the fine structure in the potential.  
As is well known, $V_{\phi\phi}$ represents the effective mass of the
scalar field.  In this sense, it is considered that the source
function $g$ traces the change of the mass of the inflaton.  Of
course, if the inflaton is coupled with other massive scalar fields
during inflation, then the effective mass of the inflaton can be
affected by the evolution of such massive fields.  Hence, such kind of
fine structures in the primordial power spectrum may provide some
clues for high energy
physics~\cite{Romano:2008rr,Barnaby:2009dd,Barnaby:2009mc,Green:2009ds,Achucarro:2010da,Jackson:2011qg}.

Let us now focus on the width of the source function $g$ 
which may control the width of the primordial power spectrum. 
On the contrary, the width of $g$ should relate with the one of the inflaton potential. 
Figure. \ref{fig:width compare} shows the relation of the $\Delta\phi_0$ 
to the parameter $\Delta\phi_g$ which is defined as the width of $g(\phi)$ as is shown in 
figure \ref{fig:source and pote}.  From this figure, it is clear that
the width of $g(\phi)$ is a monotonic function of the one of the
inflaton potential.  Hence, when we can find the form of the function
$g$ through the observations of the primordial power spectrum, we can
easily reconstruct the potential form of the inflaton.  In the
following discussion, we consider the effect of the anomalous
structure of the inflaton potential on the primordial power spectrum
with changing the functional form of $g$.

Before closing this subsection, 
we would like to remark that we can separate $g$ into two components, namely
the smooth part $g_{\rm GSR}$ whose integration in terms of time 
produces a power law component in the primordial power spectrum 
and the gap part $g_{\rm gap}$ which corresponds to a fine structure in the spectrum as 
\begin{equation}
  g=g_\mathrm{GSR}+g_\mathrm{gap}.
\end{equation}
Note that the function $f(x)$ in 
eq.~(\ref{eq:analytical formula of power spectrum}) can be derived by
integrating the function $g(x)$ from its definition eq.~(\ref{eq:definition of source}).
Here the power law part can be written by using the spectral index $n_s$, which is defined as 
${\cal P}_{\cal R} \propto k^{n_s-1}$, as 
\begin{equation}
  g_\mathrm{GSR}=\frac{n_s-1}{2}\frac{n_s-3}{2}-(n_s-1)~. 
\end{equation}
In this paper, we adopt $n_s=0.96$\cite{Komatsu:2010fb}. 

On the other hand, we employ a general form of the inflaton potential with steps and bumps
to investigate whether we can obtain the observed fine structure at $\ell\approx
100\sim140$ or not in the next subsection.  
It turns out that the resultant functional form of $g_{\rm gap}$ can be described by 
the differential Gaussian function.  

\subsection{Fine structures of the power spectrum controlled by $g$}

\begin{figure}[t]
  \begin{tabular}{cc}
    \begin{minipage}{0.5\textwidth}
      \begin{center}
        \includegraphics[width=\textwidth]
        {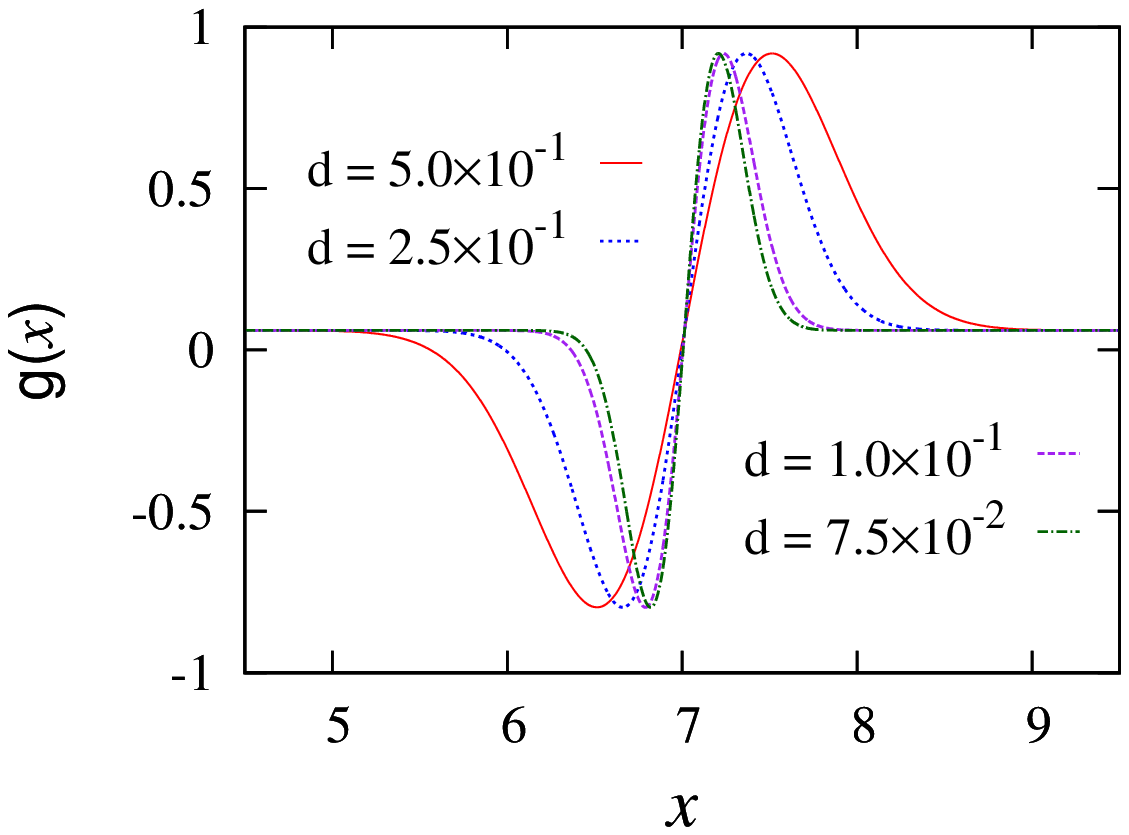}
      \end{center}
    \end{minipage}
    \begin{minipage}{0.5\textwidth}
      \begin{center}
        \includegraphics[width=\textwidth]
        {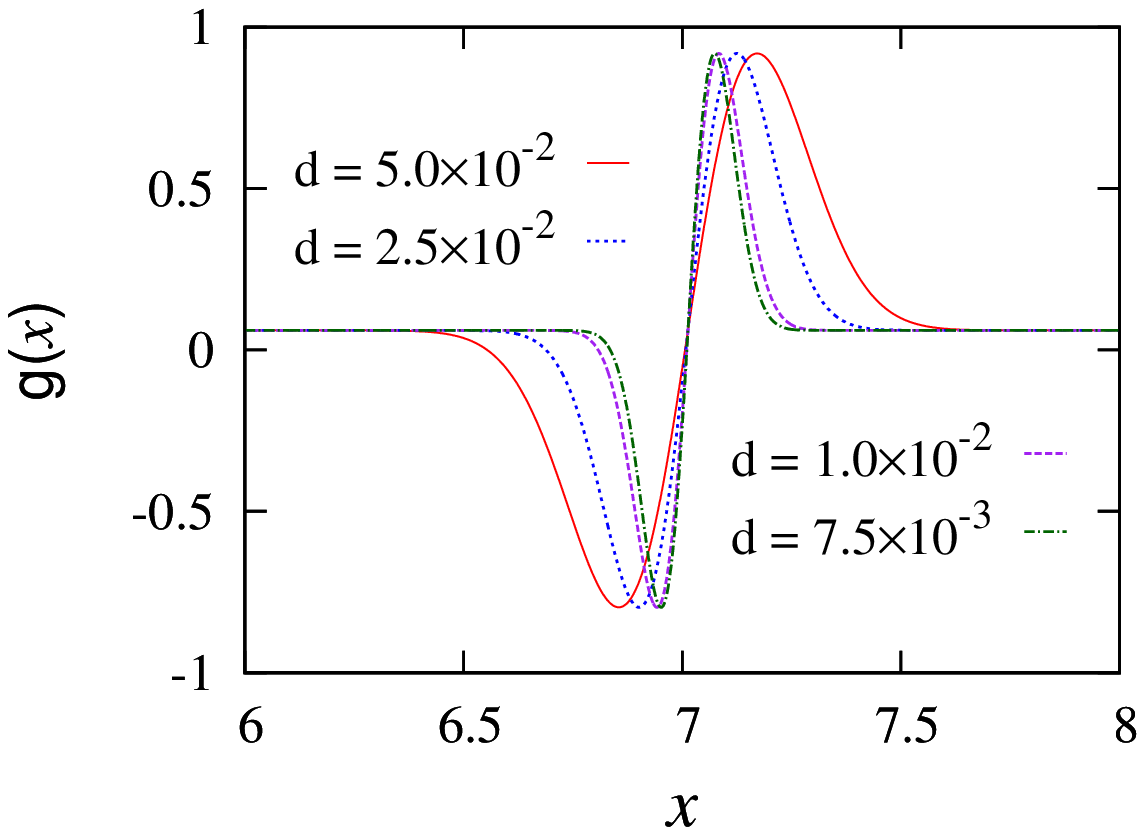}
      \end{center}
    \end{minipage}
  \end{tabular}
  \caption{The source function $g$ for the differential Gaussian model with $n=1$. 
    We set the parameter $d$ from 
    $5.0\times 10^{-1}$ to $7.5\times 10^{-2}$ for the left panel and from 
    $5.0\times 10^{-2}$ to $7.5\times 10^{-3}$ for the right panel. }
  \label{fig:differential gaussian g}
\end{figure}
\begin{figure}
  \begin{tabular}{cc}
    \begin{minipage}{0.5\textwidth}
      \begin{center}
        \includegraphics[width=\textwidth]
        {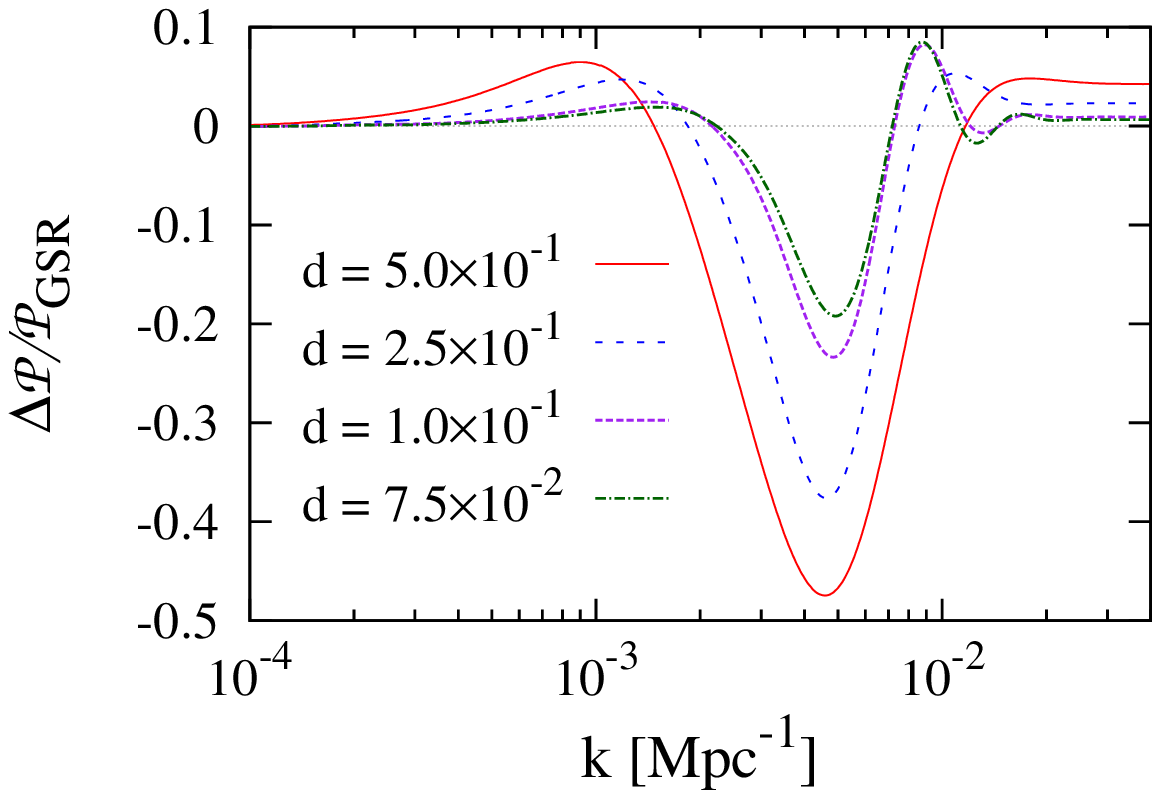}
      \end{center}
    \end{minipage}
    \begin{minipage}{0.5\textwidth}
      \begin{center}
        \includegraphics[width=\textwidth]
        {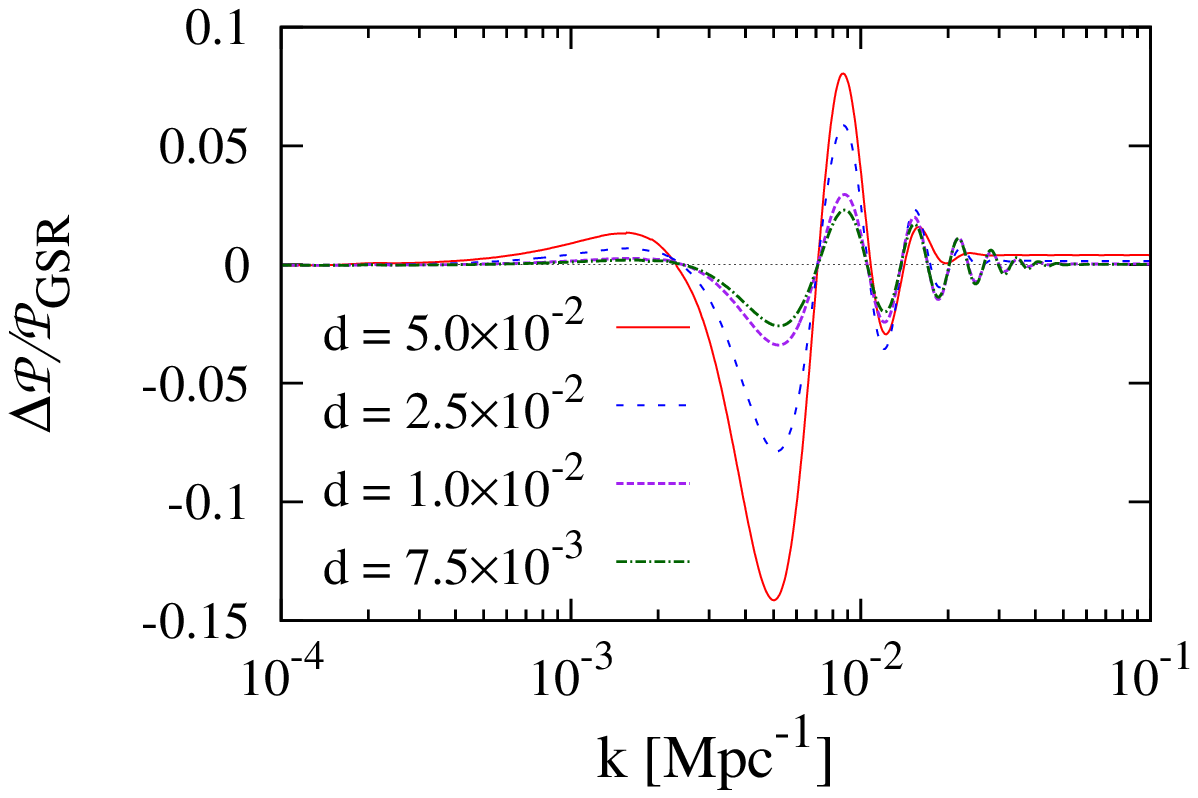}
      \end{center}
    \end{minipage}
  \end{tabular}
  \caption{Ratio between $\Delta {\cal P}(k) \equiv {\cal P}_{\cal
   R}(k) - {\cal P}_\mathrm{GSR}(k)$ and ${\cal P}_\mathrm{GSR}(k)$,
   where $ {\cal P}_{\cal R}(k)$ is the full power spectrum and ${\cal
   P}_\mathrm{GSR}(k)$ is the power-law power spectrum which is
   constructed from $g_\mathrm{GSR}$. As $d$, which is the width of
   $g$, decreases, the width of the structure in the primordial power
   spectrum also decreases as is shown in the left panel.  If $d \leq
   1.0\time 10^{-1}$, however, the width of the structure no longer
   decreases.}
  \label{fig:differential gaussian power}
\end{figure}

In order to investigate the fine structure of the primordial power spectrum
due to the modification of the slow-roll dynamics, 
we employ the differential Gaussian function as 
the gap part of the source function $g_\mathrm{gap}(x)$, given by
\begin{eqnarray}
  g_\mathrm{gap}(x)=A\,\exp{\left[-\frac{(x-b)^2}{d}\right]}\sum_{i=0}^{[n/2]}
  \frac{n!(-d)^{i-n}}{i!(n-2i)!}\left\{2(x-b)\right\}^{n-2i},~{\rm for}~n=1,2,\cdots~.
  \label{eq:nth order differential gaussian}
\end{eqnarray}
Here $[n/2]$ is the Gauss symbol.  
It should be noticed that $n=1$ and $n=2$ correspond to a step and a
bump / dip in the inflaton potential, respectively.  
Therefore the differential Gaussian function for $g_\mathrm{gap}$
is a natural and general extension of 
the inflaton potential with simple fine features.  
Moreover, the width and height of the features can be easily controlled by 
changing the parameters  $d$ and $A$, respectively.  

Our goal is to explain the newly found fine structure at $\ell\approx 100\sim 140$.  
By employing the inversion method, as we have already mentioned, 
Ichiki et al. \cite{Ichiki:2009xs} found the structure. 
They considered several different features in the primordial power spectrum 
to match the data.  
They found that a typical width of the fine feature
is $\Delta \ln k = 0.04$ and a best-fit functional form is
$\mathrm{v}^\Lambda$ one, that is a power spectrum with 
a peak at $k=k_\ast$ and a dip at $k < k_\ast$.  

In fact, it turns out such a feature in a power spectrum can be
realized even if we take the simplest functional form, i.e., $n=1$ in
eq.~(\ref{eq:nth order differential gaussian}) with an appropriate set
of parameters.  However it is not clear whether we can satisfy 
the height and width of the power spectrum simultaneously to fit the observation. 
Therefore let us begin with the $n=1$ case.  
The source function $g_\mathrm{gap}$ can be given by
\begin{equation}
  g_\mathrm{gap}(x)=\frac{c}{\sqrt{d}}\,(x-b)
  \exp{\left[-\frac{(x-b)^2}{d}\right]}.
  \label{eq:differential gaussian g}
\end{equation}
Here, the parameters $c$ and $d$
determine the height and the width of $g_\mathrm{gap}$, respectively, 
and $b (= -\ln k_0)$ corresponds to 
the position of the fine feature on the primordial power 
spectrum.  
To fit the fine structure at $\ell\approx 100\sim 140$, 
we set $k_0=0.002\,\mathrm{[Mpc^{-1}]}$.  
To maximize the amplitude of the feature under the assumption 
$g \simeq {\cal O}(1)$, we set $c = 2$. 

In figure \ref{fig:differential gaussian g}, we show the source function $g$ for 
several different values of $d$ between $10^{-1}$ and $10^{-3}$. 
As we decrease the value of $d$, the width of $g_{\rm gap}$ becomes finer as we expected.   
\begin{figure}[t]
  \begin{tabular}{cc}
    \begin{minipage}{0.5\textwidth}
      \begin{center}
        \includegraphics[width=\textwidth]
        {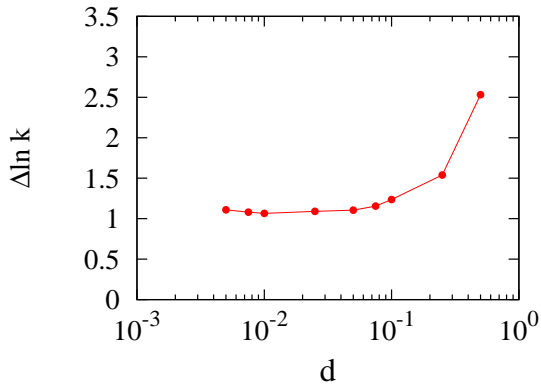}
        \caption{The width of the structure in the power spectrum in
          logarithmic space, $\Delta\ln k$ as a function of the width
          of the source function $d$ for the $n=1$ case. At $d\leq
          1.0\times 10^{-1}$, $\Delta\ln k$ becomes roughly constant.
          }
        \label{fig:d_dellogk_n=1}
      \end{center}
    \end{minipage}
    \begin{minipage}{0.5\textwidth}
      \begin{center}
        \includegraphics[width=\textwidth]
        {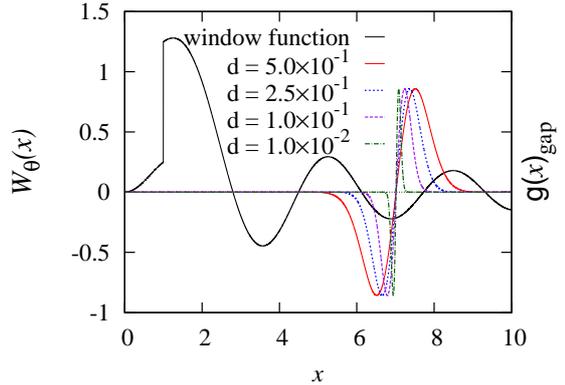}
        \caption{The window function (black line) and the source function 
          $g_\mathrm{gap}$. As the width of the primordial power spectrum features 
          does not change, the width of the $g_\mathrm{gap}$ is less than the 
          that of window function. }
        \label{fig:window_compare}
      \end{center}
    \end{minipage}
  \end{tabular}
\end{figure}

In figure~\ref{fig:differential gaussian power},
we show deviation of the primordial power spectrum ${\cal P}_{\cal R}(k)$ with $g=g_{\rm GSR} + g_{\rm gap}$ 
from the power-law power spectrum ${\cal P}_{\rm GSR}(k)$ with $g=g_{\rm GSR}$ for each value of $d$. 
Here the deviation $\Delta {\cal P}(k) \equiv {\cal P}_{\cal R}(k) - {\cal P}_{\rm GSR}(k)$.   
From this figure, it is found that the form of the power spectrum shows a single dip 
for $d \gtrsim 10^{-1}$, while there appears a dip and a bump in the power spectrum, that is 
the  $\mathrm{v}^\Lambda$ type, for $d \lesssim 10^{-1}$. 
If we push $d$ down to $d \simeq {\cal O}(10^{-2})$, the power spectrum starts to oscillate.   

The width of the power spectrum $\Delta\ln k$ is defined by the
distance between the wave numbers where $\Delta {\cal P}(k) / {\cal
P}_{\rm GSR}(k)$ becomes zero around the maximum value of $\Delta
{\cal P}(k)/{\cal P}_{\rm GSR}(k)$.  In figure~\ref{fig:d_dellogk_n=1}
, we show the dependence of the width $\Delta
\ln k$ on the parameter $d$ which controls the width of $g_{\rm gap}$.
In fact, this figure shows that $\Delta\ln k$ decreases as $d$
decreases for $d \gtrsim 1.0\times 10^{-1}$.  However, $\Delta\ln k$
starts to saturate for $ d \lesssim 1.0 \times 10^{-1}$, whose value
is about $\Delta \ln k \simeq {\cal O}(1)$.  It turns out that it is
very difficult to realize the observed width $\Delta\ln k=0.04$ in
this case.  In order to explain this result, we compare the source
function $g$ with the window function $W_\theta(x)$ in figure
\ref{fig:window_compare}.  From this figure, we see that the width of
the source function becomes smaller than the one of the window
function when $d \lesssim 1.0 \times 10^{-1}$.  Since the width of the
primordial power spectrum is determined by the products of $g$ and
$W_\theta(x)$, it is mainly determined by the width of the window
function in this case.  This causes the saturation because the window
function smoothes out the finer feature of the source function.  On
the other hand, if $d \gtrsim 1.0 \times 10^{-1}$, the width of the
source function becomes larger than the one of the window function.
Accordingly, the width of the primordial power spectrum is controlled
by the one of the source function, which is $d$.

\begin{figure}[t]
  \begin{tabular}{cc}
    \begin{minipage}{0.5\textwidth}
      \begin{center}
        \includegraphics[width=1.0\textwidth]
        {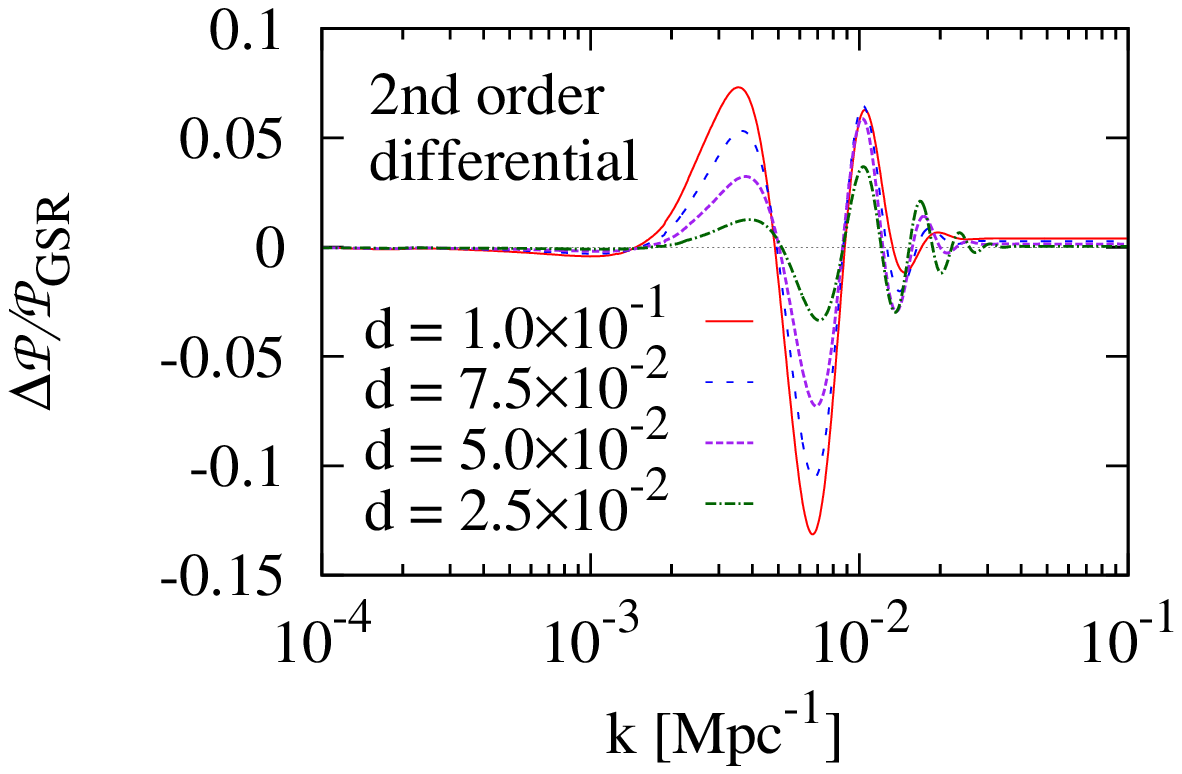}
      \end{center}
    \end{minipage}
    \begin{minipage}{0.5\textwidth}
      \begin{center}
        \includegraphics[width=\textwidth]
        {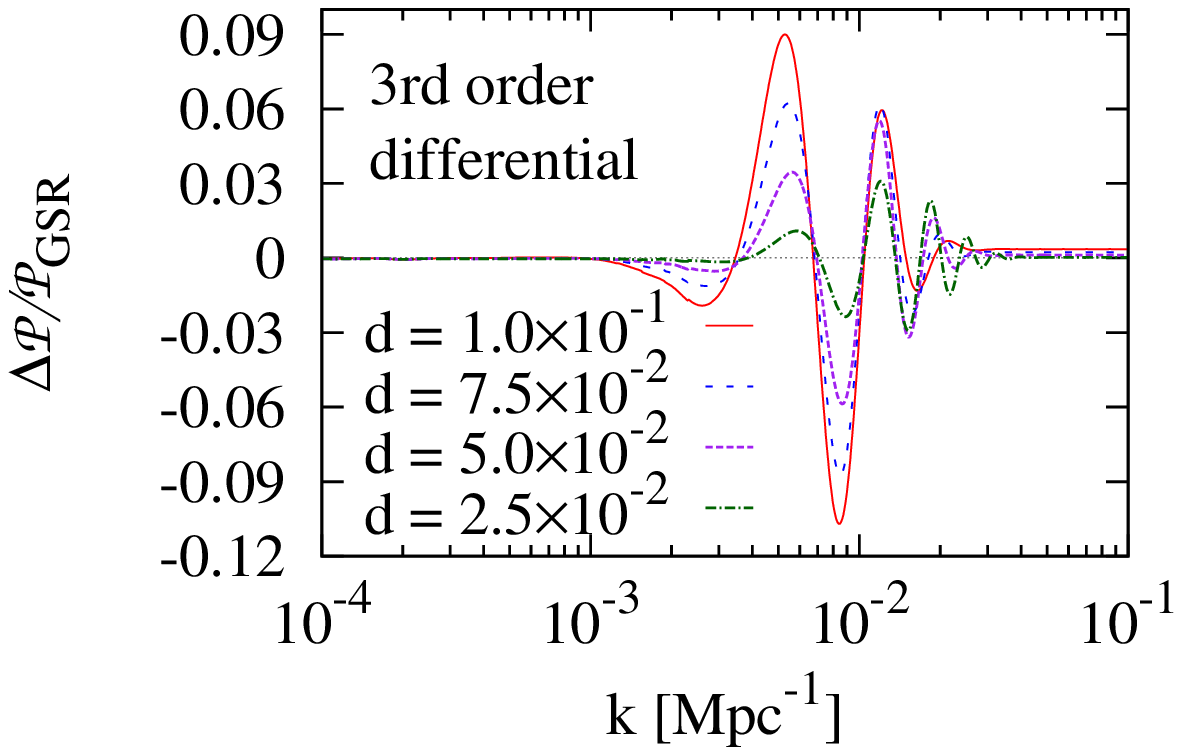}
      \end{center}
    \end{minipage}
  \end{tabular}
  \caption{Ratio between $\Delta {\cal P}(k) \equiv {\cal P}_{\cal
      R}(k) - {\cal P}_\mathrm{GSR}(k)$ and ${\cal P}_\mathrm{GSR}(k)$.
    for the $n=2$ in the left panel and for $n=3$ in the right panel.}
  \label{fig:analyze power spectrum}
\end{figure}

Now, let us investigate higher order differential Gaussian models
given by eq.~\eqref{eq:nth order differential gaussian}, i.e., models
with $n \geq 2$.  In figure~\ref{fig:analyze power spectrum}, deviations
of the full primordial power spectrum from the power-law one for $n=2$
and $3$ are shown.  We find that the width of the structure becomes
finer for higher order models, although the saturation for $d \leq
1\times 10^{-1}$ can be seen as the model with $n=1$.  
In figure~\ref{fig:d_dellogk_n=2,3}, the widths of the structure $\Delta\ln
k$ are plotted as a function of $d$ for $n=1, 2$ and $3$ to see this tendency 
more clearly.  

We can conclude that, to have a finer structure in the primordial
power spectrum, we need to employ a higher order differential Gaussian
function as a source function.  It is because a higher order
differential Gaussian function has more numbers of oscillation for a
given $d$, and these oscillations pick up the larger wave number part
of the window function, which actually produces finer structure in the
power spectrum.  

\begin{figure}[t]
  \begin{tabular}{cc}
    \begin{minipage}{0.5\textwidth}
      \begin{center}
        \includegraphics[width=1.0\textwidth]
        {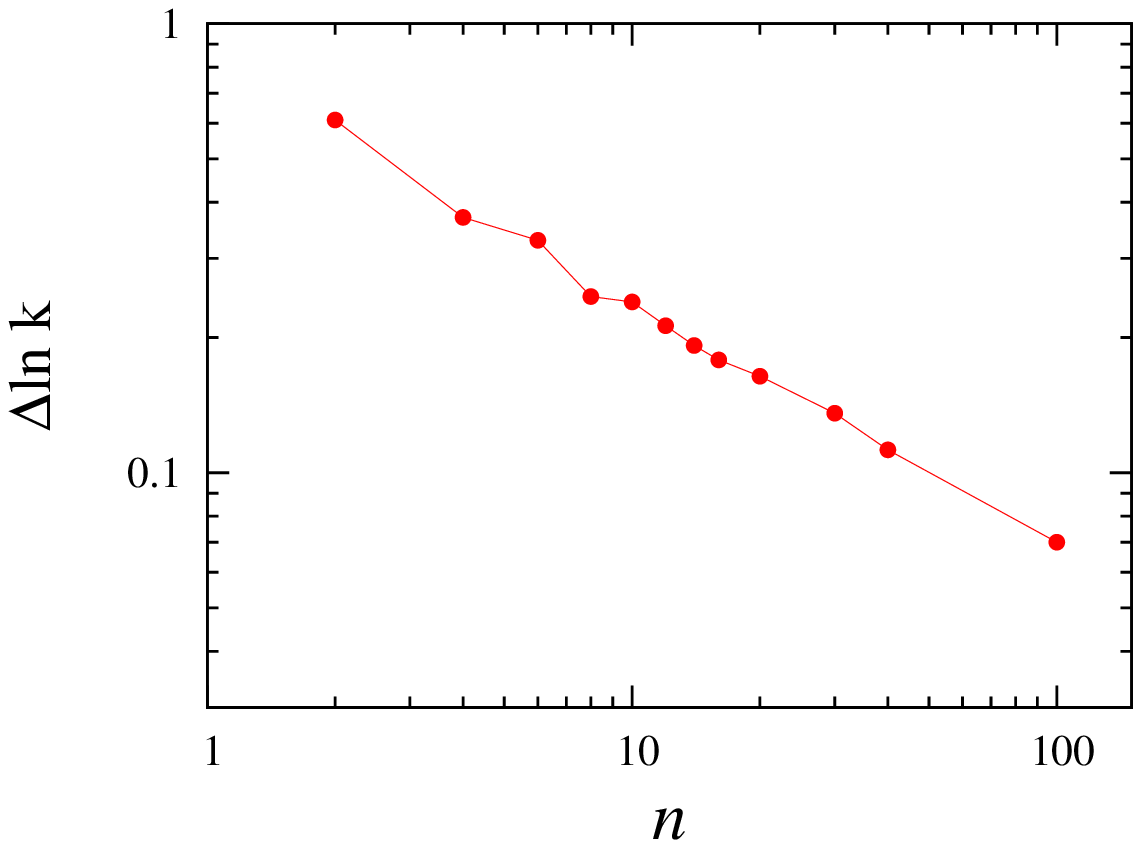}
      \end{center}
    \end{minipage}
    \begin{minipage}{0.5\textwidth}
      \begin{center}
        \includegraphics[width=\textwidth]
        {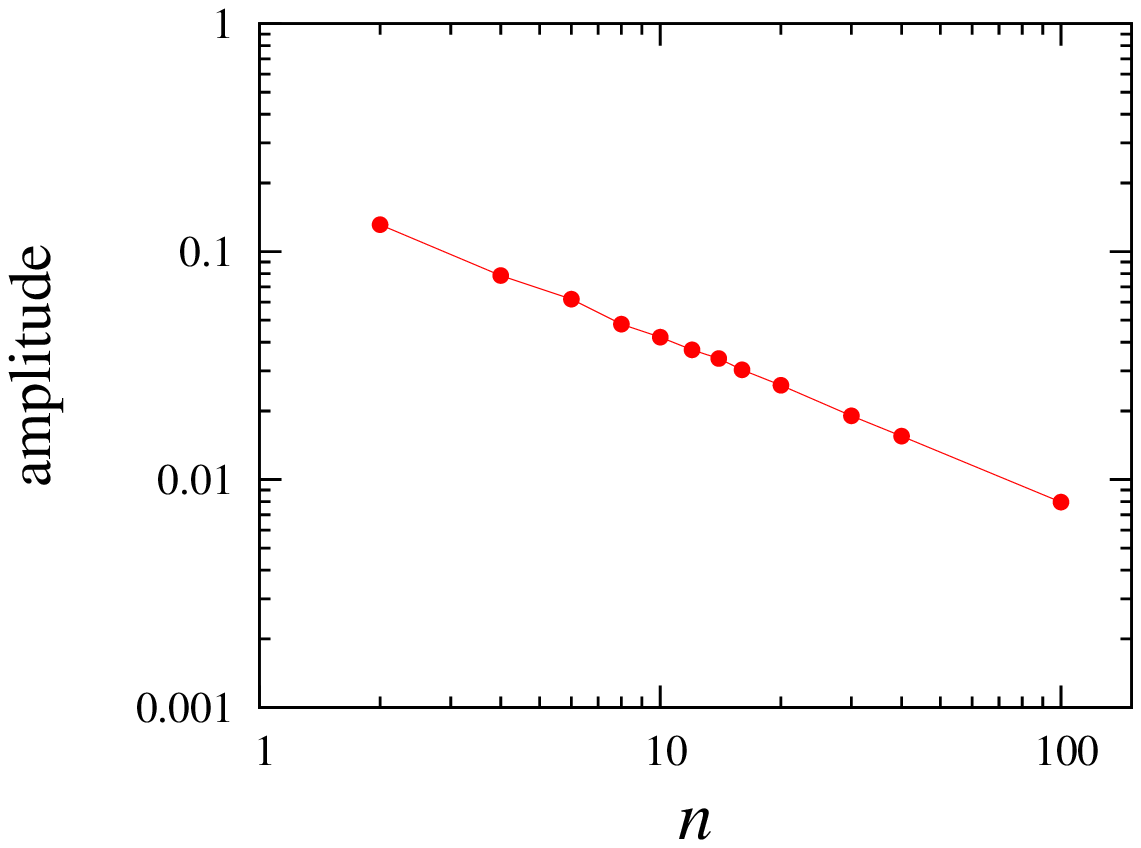}
      \end{center}
    \end{minipage}
  \end{tabular}
  \caption{The width of the fine structure $\Delta\ln k$ as the function of 
    the number of differentiation $n$ in the left panel.
    Even if the $n$ is increased to $100$, $\Delta\ln k$
    can not reach at the observed value $\Delta\ln k=0.04$. 
    And the right panel shows that the maximum amplitude 
    of $\Delta{\cal P}(k)/{\cal P}_{\mathrm{GSR}}(k)$ as the function of 
    $n$. 
    The amplitude drops below $10^{-2}$ for $n =100$, which is too small 
    to explain the observed structure.}
   \label{fig:conclusion of nth order differential Gaussian}
\end{figure}

In figure~\ref{fig:conclusion of nth order differential Gaussian}, 
we push the number of differentiation $n$ much further up to $100$.    
We can almost reach the observed value $\Delta \ln k = 0.04$.  
However, there is a caveat.  If we take $n$ to be a large number, 
the maximum value of $\Delta {\cal P}/{\cal P}_{\rm GSR}$, which 
we refere as the amplitude of the structure hereafter, becomes 
smaller.  As is shown in the right panel of 
figure~\ref{fig:conclusion of nth order differential Gaussian}, 
the amplitude drops below $10^{-2}$ for $n =100$, which is way 
too small to explain the observed structure.  

\begin{figure}[t]
  \begin{center}
    \includegraphics[width=0.5\textwidth]
    {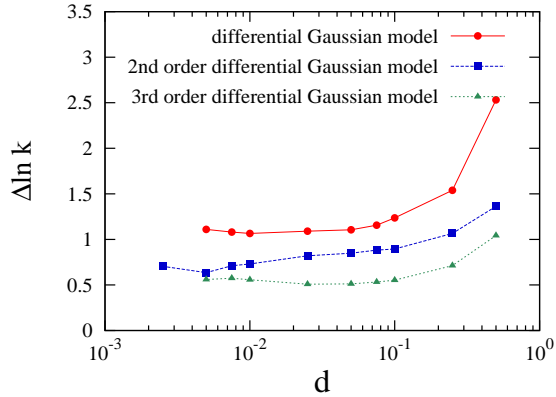}
  \end{center}
  \caption{The width of the structure in the power spectrum in
      logarithmic space, $\Delta\ln k$ as a function of the width
      of the source function $d$ for the $n=2$ and $3$ cases  
      of differential Gaussian models with the $n=1$
      one (figure~\ref{fig:d_dellogk_n=1}).  It is shown that
      $\Delta\ln k$ becomes smaller for larger $n$ models.}
  \label{fig:d_dellogk_n=2,3}
\end{figure}

\section{Summary and discussion}
In this paper, we have discussed the fine features on the primordial
power spectrum which can not be characterized by a simple power-law
spectrum.  Previous works to focus on the fine structure of the CMB
temperature power spectrum at the multipole $l=20\sim40$ have shown
that such an oscillating feature can be realized by the modification
of the slow-roll dynamics during inflation.  Based on this idea, we
have examined the possibility of explaining a newly found finer
feature of the primordial power spectrum which are reconstructed from
WMAP data corresponding to the multipole $\ell=100\sim 140$ by using
the inversion method by ref.~\cite{Ichiki:2009xs}.  

Employing the general slow-roll formula presented by
ref.~\cite{Stewart:2001cd}, we can characterize the power spectrum by
the window function, which is a oscillating function with a sharp cut
off corresponding to the horizon crossing scale, and the source
function, which is described by the slow roll parameters and their
time derivatives.   For the source function, we have used 
the differential Gaussian function, since it can naturally describe 
the structures such as step, dip and bump in the inflaton potential.  
Moreover, we can easily control the width of the structure and the 
number of oscillations of the source function.  

We have found that the width of the fine structure $\Delta\ln k$ 
is controlled by the width of the window function in the first place
while it also depends on the width of the source function $d$. 
In fact, since the window function smoothes out the structure of the source 
function, $\Delta\ln k$ saturates when we decrease the value of $d$ 
below $ \sim 1.0 \times 10^{-1}$.
However, we have obtained the finer structure by employing the higher order 
differential Gaussian model as the source function.
It is because that oscillations of the source function pick up the 
larger wave number part which corresponds to the inner horizon part  
of the window function. 
On the other hand, the amplitude of the fine structure, which is defined by 
the maximum value of $\Delta{\cal P}(k)/{\cal
P}_\mathrm{GSR}(k)$, becomes smaller as we increase the index of differentiation 
$n$ of the source function.  
It turns out that it is hardly possible to recover the observed fine structure 
at $\ell=100\sim 140$ with simultaneously satisfying the width and the amplitude.  

There might be a possible solution to expalin the fine structure with
introducing a higher order correction of the general slow-roll
formula~\cite{Choe:2004zg}, which allows us to have a large value of
the source function $\vert g\vert \gtrsim 1$.  In this case, the
source function may dominate the window function and control the fine
structure of the power spectrum.  Such a source function may generate
a structure with a large amplitude.  However, Dvorkin and
Hu~\cite{Dvorkin:2009ne} studied the effect of the second order term
and found that only a little enhancement of the amplitude takes place.

The only way to explain the observed fine structure might be 
to have a fine tuned source function which perfectly cancels the 
oscillations of the window function.  Of cource, such a source 
function is compleltely ad-hoc.
We may conclude that, therefore, it is almost impossible to 
generate the observed fine structure from a simple modification of 
the slow roll dynamics by adding specific features in the inflaton potential. 

\acknowledgments
We would like to thank Keitaro Takahashi and Kiyotomo Ichiki for
useful discussion. This work is supported by JSPS Grand-in-Aid for Scientific 
Research under Grant Nos.22340056. This work is supported in part by the 
Grand-in-Aid for Scientific Research on Priority Areas No. 467 "Probing the 
Dark Energy through an Extremely Wide and Deep Survey with Subaru Telescope"
and by the Grant-in-Aid for Nagoya University Global COE Program, "Quest for 
Fundamental Principles in the Universe: from Particles to the Solar System and 
the Cosmos," from the Ministry of Education, Culture, Sports, Science and 
Technology of Japan.  This research has also been supported in part by World Premier
International Research Center Initiative, MEXT, Japan.

\appendix
\section{Validity of general slow-roll formula}
\label{sec:valid}
In this appendix, we check
the validity of the analytical formula
given by eq.~\eqref{eq:analytical formula of power spectrum},
by comparing the approximate solution with a numerical calculation.
Here we employ the chaotic inflation model with
a dip in the potential, whose potential is given by eq.~(\ref{eq:bumppotential})
with $c < 0$.

\begin{figure}
 \begin{tabular}{cc}
   \begin{minipage}{0.5\textwidth}
     \begin{center}
       \includegraphics[width=\textwidth]
       {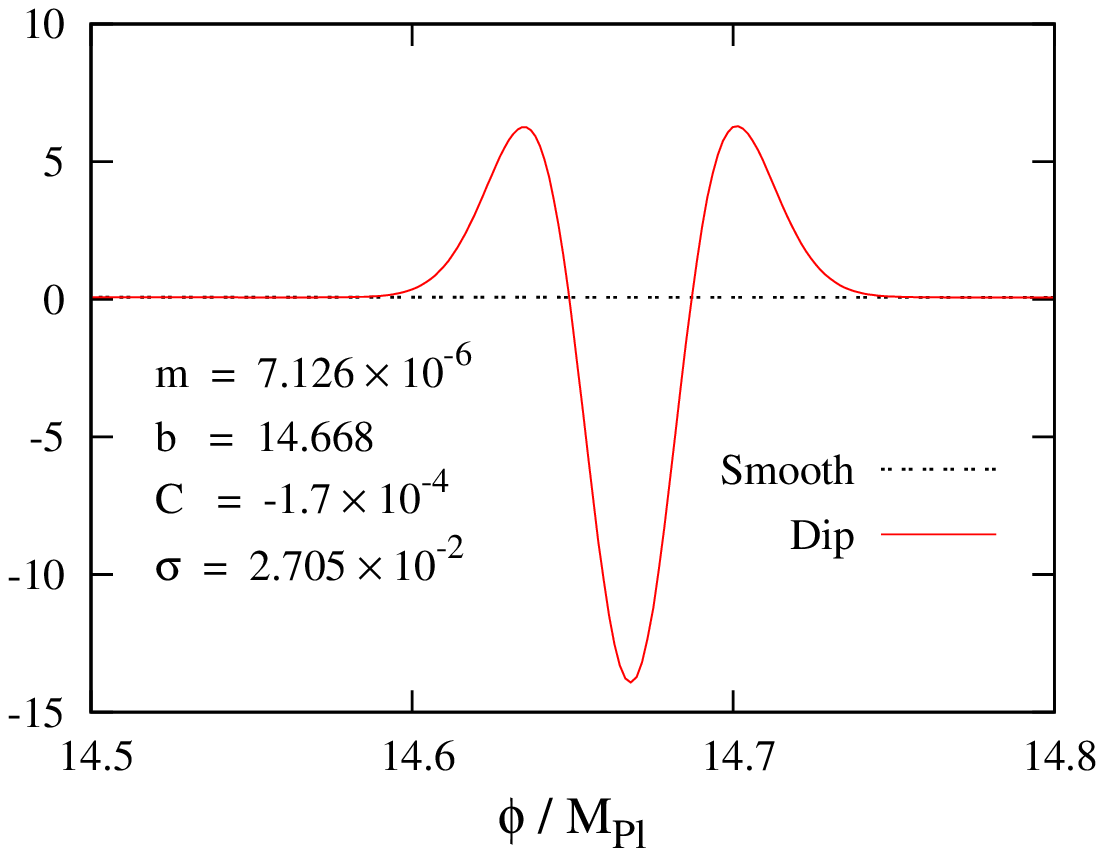}
     \end{center}
   \end{minipage}
   \begin{minipage}{0.5\textwidth}
     \begin{center}
       \includegraphics[width=0.7\textwidth,angle=-90]
       {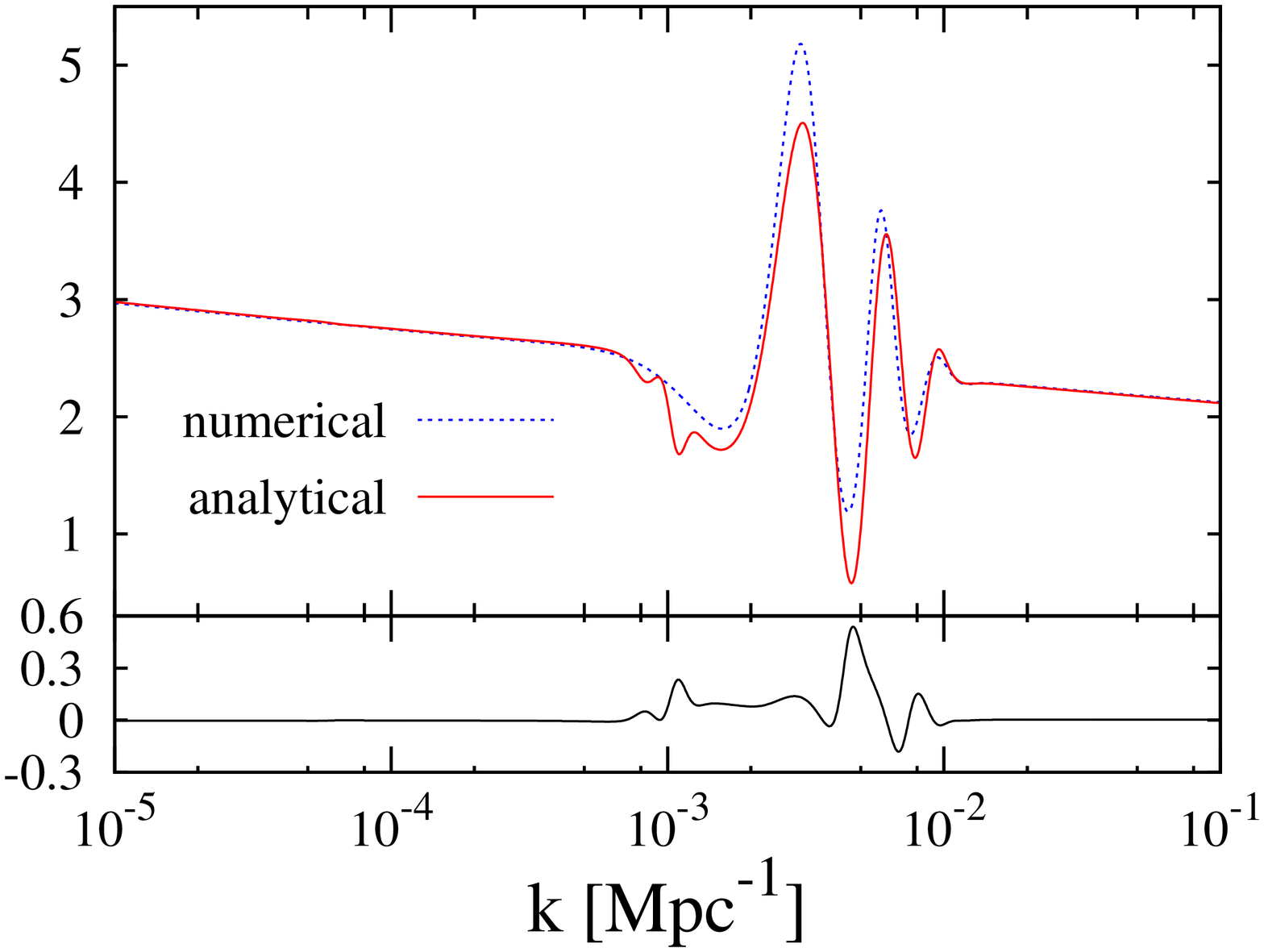}
     \end{center}
   \end{minipage}
 \end{tabular}
\caption{The source functions and primordial power spectra for the dip model
are shown in the left panels and right panels, respectively.
 We set the model parameter $c$ to $- 5.6\times 10^{-5}$,
  $c = - 5.6\times 10^{-4}$ and
    $c = - 1.7\times 10^{-4}$ from top to bottom.}
 \label{fig:dip_validity}
\end{figure}

In the left panels of figure \ref{fig:dip_validity},
we show the source function $g$ with $m = 7.0 \times 10^{-6}$,
$\phi_0 = 14.67$ and  $\Delta \phi_0^2 = 2.7 \times 10^{-2}$ for
$c = -5.6 \times 10^{-5}$, $-5.6 \times 10^{-4}$ and $-1.7 \times 10^{-4}$ from
top to bottom.
These values of model parameters correspond to the ones employed in
ref. \cite{Mortonson:2009qv}, in which the feature in the WMAP data at
$\ell =20\sim 40$ are fitted.
Note that we have made a conversion of parameters since they assumed
a small step in the potential instead of a dip with keeping the width
and the position of the feature.
From these panels,
we can see that the maximum value of the source function $g$
becomes larger as the absolute value of $c$ increases.
In the right panels of figure~\ref{fig:dip_validity},
we show the primordial power spectra
of respective models
obtained by making use of
the general slow-roll formula (red solid line) and the numerical
calculation (blue dotted line)
and also the fractional errors between the two results.

\begin{figure}[t]
 \begin{tabular}{cc}
   \begin{minipage}{0.5\textwidth}
     \begin{center}
       \includegraphics[width=\textwidth ,angle = 0]
       {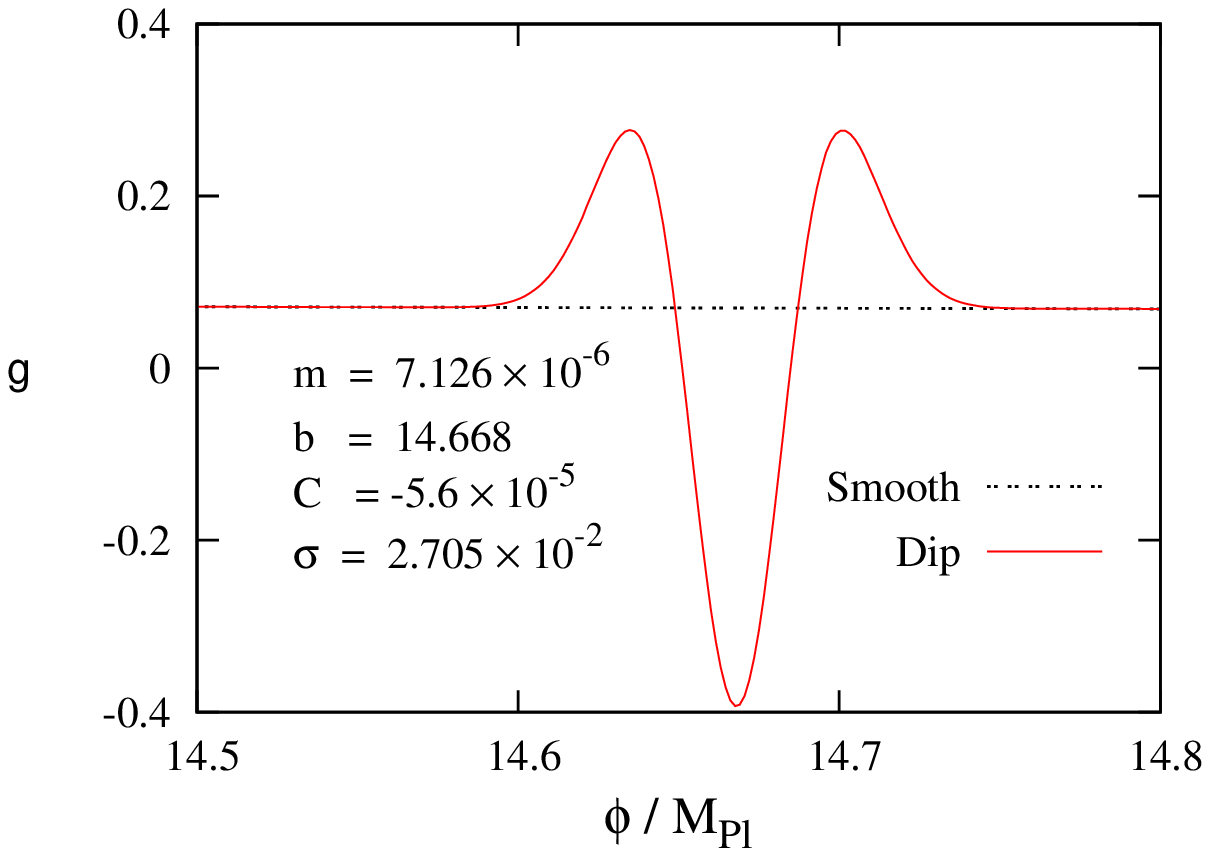}
     \end{center}
   \end{minipage}
   \begin{minipage}{0.5\textwidth}
     \begin{center}
       \includegraphics[width=0.7\textwidth ,angle = -90]
       {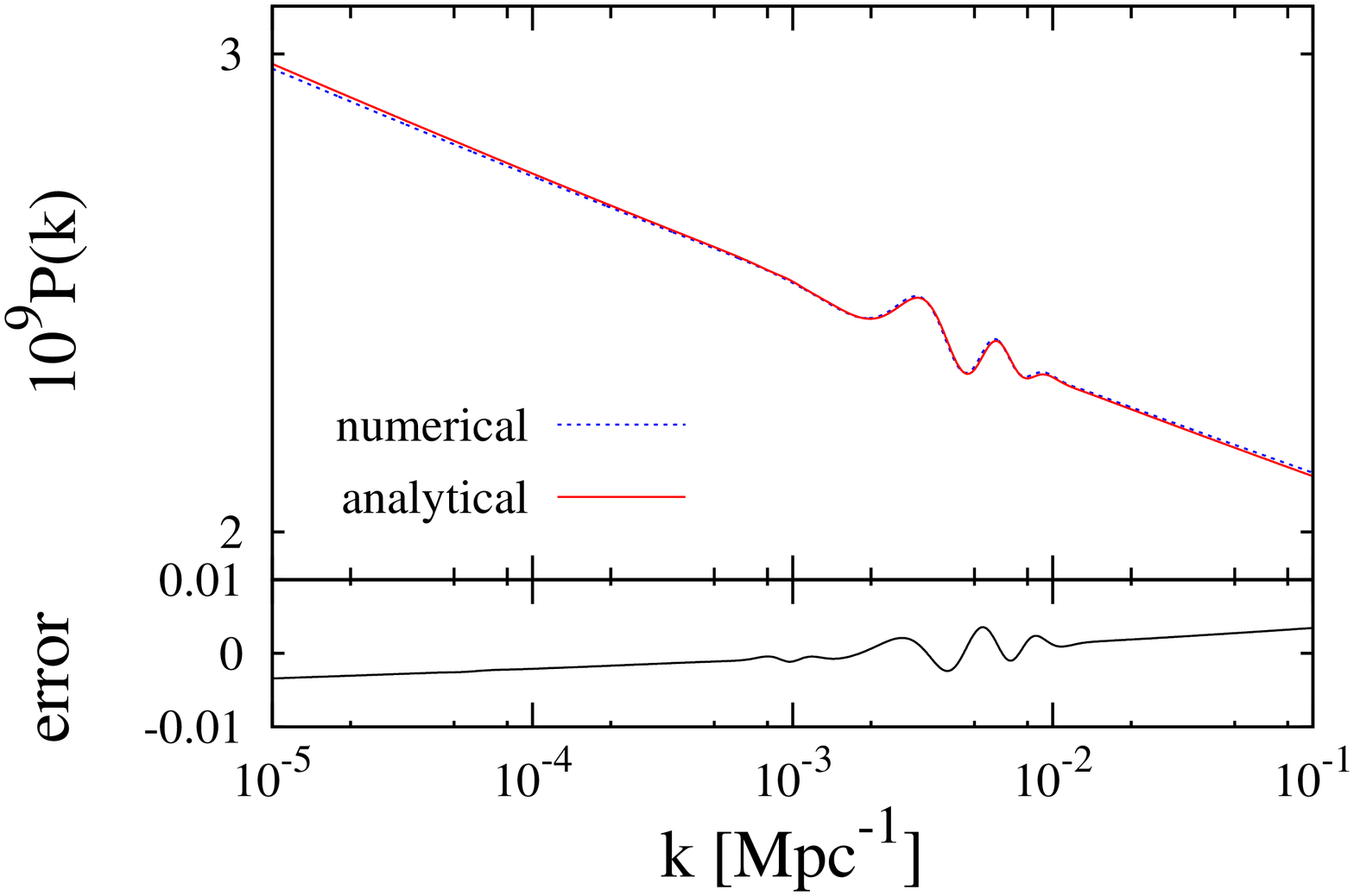}
     \end{center}
   \end{minipage}
   \label{fig:source dip}
 \end{tabular}
\end{figure}
\begin{figure}
 \begin{tabular}{cc}
   \begin{minipage}{0.5\textwidth}
     \begin{center}
       \includegraphics[width=\textwidth]
       {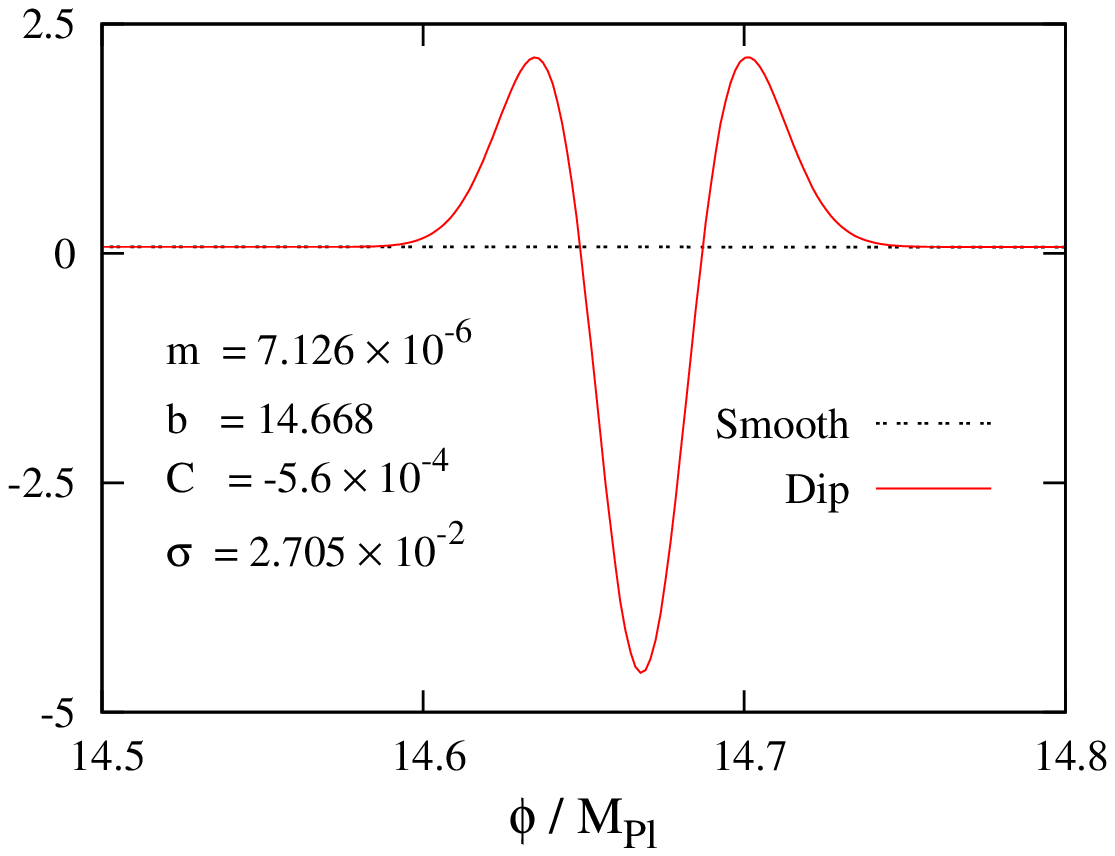}
     \end{center}
   \end{minipage}
   \begin{minipage}{0.5\textwidth}
     \begin{center}
       \includegraphics[width=0.7\textwidth ,angle = -90]
       {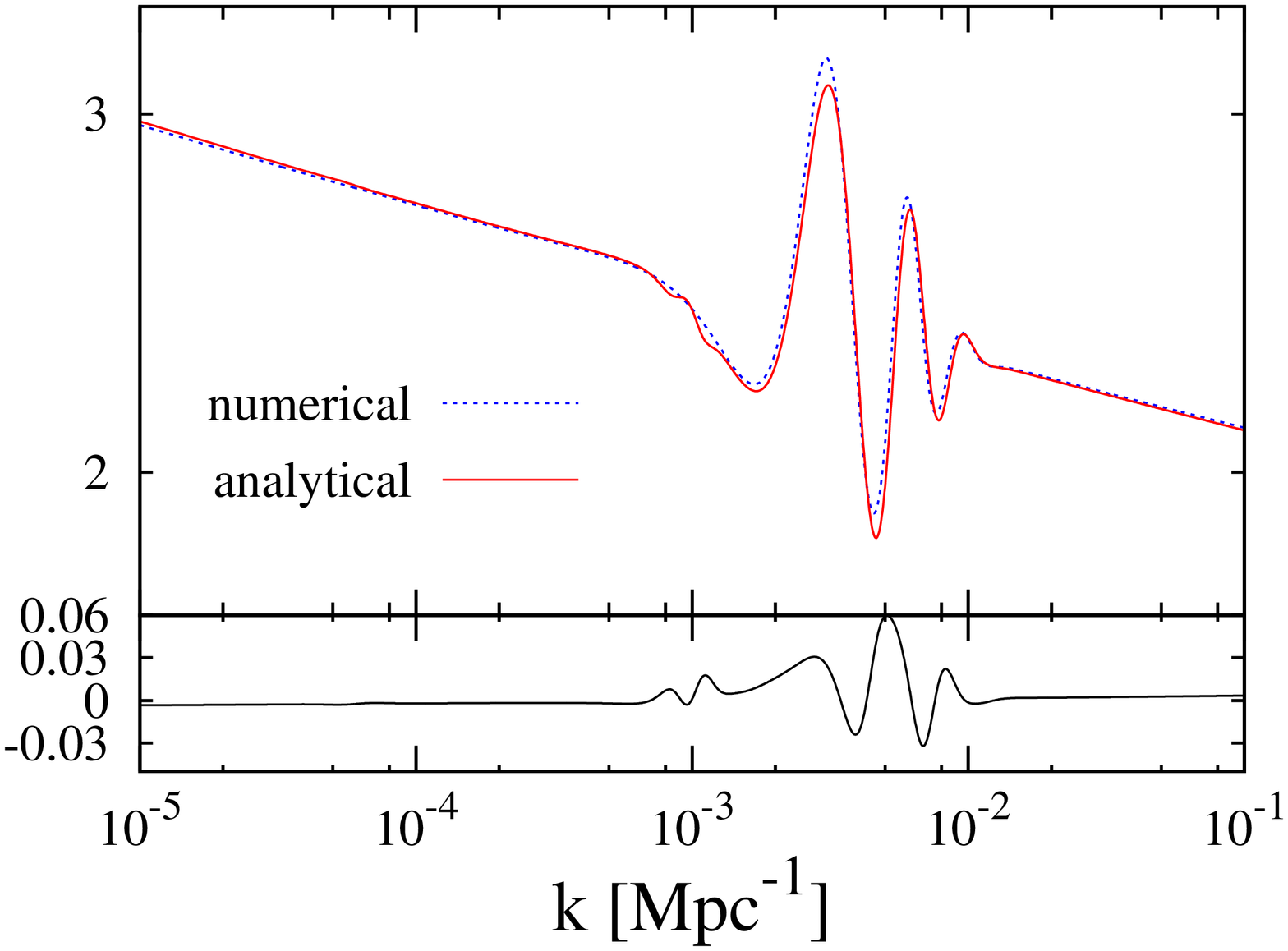}
     \end{center}
   \end{minipage}
 \end{tabular}
\end{figure}

From this figure,
we find that
around the structure of the primordial power spectrum
the fractional error between
the approximate solution and the exact solution obtained
by the numerical calculation becomes larger as the parameter $c$
increases.
If $10\%$ errors are allowed,
it is confirmed that the general slow-roll approximation
is valid for the cases with $g = {\cal O}(1)$ around the
maximum such as a case with $c = -5.6 \times 10^{-4}$ for the dip model.
Therefore,
in the context we have employed the cases where the maximum value of
the source function $g$ is order of unity.

\end{document}